\documentclass[%
 aip,pop
 sd,%
 amsmath,amssymb,
 reprint,%
]{revtex4-1}

\usepackage{graphicx}   
\usepackage{color}      
\usepackage{mathrsfs}   
\usepackage[colorinlistoftodos]{todonotes} 
\usepackage{eurosym}
\usepackage{dcolumn}
\usepackage{bm}
\usepackage{multirow}

\expandafter\let\csname equation*\endcsname\relax
\expandafter\let\csname endequation*\endcsname\relax
\usepackage{amsmath}

\begin{document}

\preprint{AIP/123-QED}

\title[Alternative propellants for HET: Insight from 2D PIC simulations]{The effect of alternative propellants on the electron drift instability in Hall-effect thrusters: Insight from 2D Particle-In-Cell simulations}

\author{Vivien Croes}
\author{Antoine Tavant}
\affiliation{Laboratoire de Physique des Plasmas, CNRS, Sorbonne Universit\'e, Universit\'e Paris Sud, \'Ecole Polytechnique, F-91120 Palaiseau, France}
\affiliation{Safran Aircraft Engines, Electric Propulsion Unit, F-27208 Vernon, France}
\author{Romain Lucken}
\email{romain.lucken@lpp.polytechnique.fr}
\author{Roberto Martorelli}
\affiliation{Laboratoire de Physique des Plasmas, CNRS, Sorbonne Universit\'e, Universit\'e Paris Sud, \'Ecole Polytechnique, F-91120 Palaiseau, France}
\author{Trevor Lafleur}
\affiliation{PlasmaPotential - Physics Consulting and Research, Canberra ACT 2601, Australia}
\author{Anne Bourdon}
\author{Pascal Chabert}
\affiliation{Laboratoire de Physique des Plasmas, CNRS, Sorbonne Universit\'e, Universit\'e Paris Sud, \'Ecole Polytechnique, F-91120 Palaiseau, France}

\date{\today}


\begin{abstract}
Hall-effect thrusters (HETs) operated with xenon are one of the most commonly used electric propulsion technologies for a wide range of space missions, including drag compensation in low Earth orbit, station-keeping, and orbital insertion, as access to space becomes more affordable. Although anomalous electron transport, the electron drift instability (EDI), and secondary electron emission (SEE) have been studied experimentally and numerically in xenon-based HETs, the impact of alternative propellants is still poorly characterized. In this work, a two-dimensional particle-in-cell/Monte Carlo collision (PIC/MCC) code is used to model the $(r-\theta)$ plane of a HET operated separately with four different noble gases: xenon, krypton, argon, and helium. Models for electron induced secondary electron emission (SEE) and dielectric walls are implemented in order to investigate the coupling between the propellant choice and the radial thruster walls. For all conditions and propellants studied, an EDI and enhanced electron cross-field transport are observed. The frequency of the instability, as well as the electron mobility, are compared with analytical expressions from a recently developed kinetic theory. Confirming this theory, it is shown that while the frequency of the EDI depends on the propellant mass, the electron mobility appears to be almost independent of the propellant choice. 

\end{abstract}

%


\maketitle
%
%
%
%



\section{Introduction}
\label{sec:intro}

\begin{figure}[!tbp]
   \centering
   \includegraphics[width=0.85\linewidth]{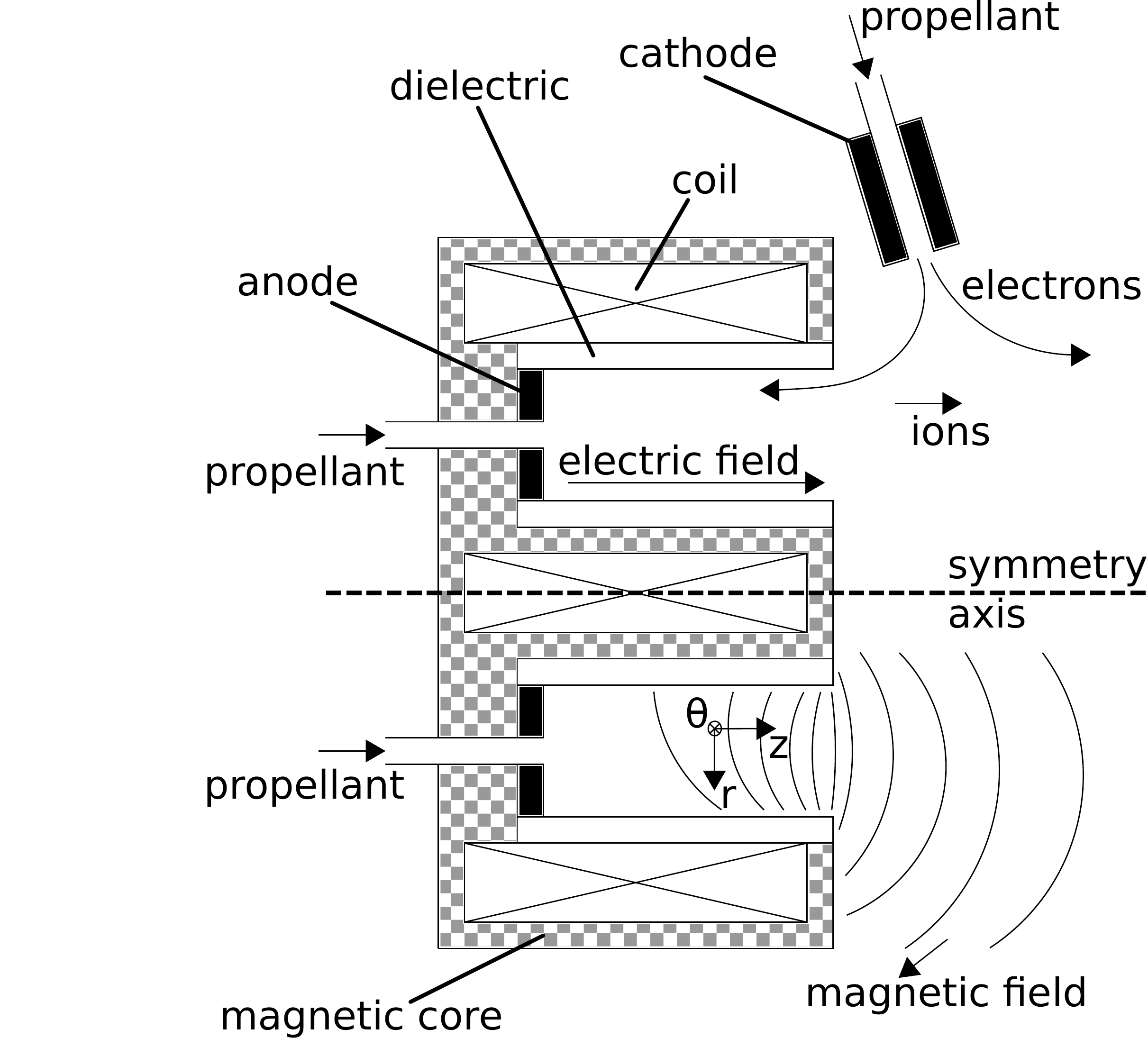}
   \caption{Schematic of a typical HET.\cite{nous}}
   \label{fig:HET_scheme}
\end{figure}

Hall-effect thrusters (HETs) are one of the most successful technologies for electric space propulsion, and are increasingly being used on commercial, military, and scientific spacecrafts.\cite{business} As illustrated in Figure \ref{fig:HET_scheme}, a typical HET consists of three main parts: \cite{Goebel,USpatent}
\begin{enumerate}
  \item An annular ceramic channel (in gray checkerboard in Figure \ref{fig:HET_scheme}) where the propellant gas is injected at the base, ionized, and accelerated. This channel has a length and a width of the order of centimeters \cite{geometry}. The density in the channel is typically in the range of $10^{17}$ to $10^{18} \, \textnormal{m}^{-3}$ for the plasma, and $10^{18}$ to $10^{20} \, \textnormal{m}^{-3}$ for the neutral gas. \cite{Trevor5}

	\item An electric circuit composed of an anode located at the base of the channel and an external hollow cathode. A large potential difference ($100$'s of volts) is applied between the anode and the cathode, which accelerates the ions to high velocities, generating thrust. The cathode provides electrons that both sustain the plasma discharge through ionization, and neutralize the ion beam. 

	\item A magnetic circuit that imposes a predominantly radial magnetic field ($10$'s of mT) near the outlet of the channel. This magnetic field enhances the residence time of the electrons in the channel and thus increases the ionization efficiency. \cite{Goebel}
\end{enumerate}

HETs are predominantly operated using xenon, due to its large mass, relatively low ionization threshold, and its chemical inertness. \cite{Goebel} However, some alternative propellants for HETs are potential substitutes to xenon\cite{Mazouffre}:

\begin{description}

	\item [Argon] would be the most cost effective solution. However, its higher ionization energy and lower mass necessitates different thruster dimensions and magnetic field topologies. Double stage HETs could be one possible solution, but so far limited success has been obtained. \cite{Anne01,2stage2,2stages}
    
    \item [Krypton] has also been tested as its relevant properties are closer to xenon, and so fewer modifications of a HET design are needed. Krypton HETs have a high specific impulse, but a lower overall efficiency than xenon. \cite{Krypton}
    
    \item [Bismuth] could also be a potential candidate. However, it has a very high melting point of about $271^{\circ}\mathrm{C}$, and could deposit on satellite surfaces. Consequently, it is not considered in the present work.
    
    \item [Iodine] is the halogen that is adjacent to xenon in the periodic table of elements, so its physical properties match the requirements of current HET designs. \cite{I2-prop, I2-exp} \textit{Busek Co. Inc.} offers both HET and gridded ion thrusters operated with iodine. Studies are ongoing to better understand the operation of electric propulsion systems using iodine. \cite{I2} However, its complex chemistry makes the proper simulation of an iodine propelled HET challenging and computationally costly, and it is thus not considered in the followings.
    
    \item [Atmospheric propellants] could also be a solution, mostly for spacecraft operated at very low altitudes. Here the residual atmosphere is used as a propellant. A \textit{PPS1350-TSD} from \textit{Safran Aircraft Engines} was tested using pure nitrogen as well as a nitrogen/oxygen mixture after ignition with xenon. Both cases showed a lower propellant efficiency and a higher anode erosion rate compared with operation using xenon. \cite{air} More recently, the European Space Agency has been testing an air-breathing HET. \cite{ESA1} However, due to the complexity of the chemical processes, HETs using atmospheric propellants are also not considered in the present work.
    
    \item [Helium] is too light to provide a good thrust efficiency, but it can be used in laboratory experiments, and studying it helps to understand the physics and scaling of the discharge since its mass is much lower than xenon.

\end{description}

HETs are complex systems, and it is known that the electron mobility across the magnetic field is anomalously high compared to the mobility predicted by classical transport theories based on standard particle collisions, \cite{Trevor1, Goebel, Trevor4, Trevor5, HET-transport-measures} particularly near the thruster exit and in the near-plume region. \cite{Trevor4, Trevor5, Trevor6, Trevor7}

The role of the inner wall material on anomalous transport has been experimentally highlighted in numerous studies \cite{Goebel, Trevor8, Trevor9, Trevor18}. However, evidence suggests that electron-wall collisions and secondary electron emissions (SEEs) are not sufficient to explain the observed cross-field electron transport. \cite{Trevor10, Trevor19, Trevor20, Trevor21, Trevor22, Trevor23}
Anomalous electron transport may be due to the presence of short-wavelength instabilities in the azimuthal direction, as highlighted by both experimental and numerical studies. \cite{Trevor24, Trevor26, Trevor27, Trevor28}
Indeed, the large electron drift velocity in the azimuthal direction acts as a driving force for these instabilities \cite{Trevor11, Trevor12, Trevor29}, which result in large amplitude fluctuations in both the plasma density and the azimuthal electric field.
These instabilities have frequencies in the MHz range, wavelengths less than a millimeter, and electric field amplitudes almost as large as the axial accelerating field itself. \cite{Trevor5}

This phenomenon has been observed in xenon-based one-dimensional (1D) particle-in-cell (PIC) Monte Carlo collisions (MCC) simulations, \cite{TrevorI} and a recent kinetic theory predicts similar instability characteristics and an enhanced electron transport. \cite{TrevorII} Two-dimensional xenon-based (2D) PIC simulations have confirmed the relevance of these predictions. \cite{Trevor11, Trevor37, nous}

As xenon production is expensive and subject to supply fluctuations \cite{Xenon}, its use imposes economic limitations. Although this issue is not strictly speaking a technological limitation, it could be a significant hurdle to the use of HETs in the future. Since the cost to launch a payload into low Earth orbit (LEO) is expected to go below about 10\,k\euro/kg with the emergence of new launch services, the required satellite propellant mass also represents an important factor. Hence, studying the physical impact of alternative propellants on HETs operation is becoming a valuable topic.

More precisely, the effect of a change of propellant on the instability remains an open question. In this paper a 2D PIC/MCC model of the $(r-\theta)$ plane of a HET is simulated, with a ``fake'' axial length to study the anomalous cross-field electron transport. Dielectric radial walls with electron-induced SEE are included in the simulation model. Four noble gases (xenon, krypton, argon, and helium) are investigated, using realistic cross section data for electron-neutral and ion-neutral interactions. Results are then compared to the kinetic theory predictions in order to challenge its quality.

Section \ref{sec:model} describes the simulation model. The properties of the instability and the anomalous electron transport obtained from the simulation are compared  with results from kinetic theory in section \ref{sec:results}. Finally, the relative influence of the characteristics of the propellant on the discharge properties is discussed in section \ref{sec:discuss}.

\section{Model description}
\label{sec:model}

The results presented in this work were obtained using an independently developed 2D-3V PIC/MCC code called LPPic2D. \cite{croes_modelisation_2017, nous,croes_study_2017,lucken_edge--center_2018,lucken_global_2017} In this code, a Cartesian geometry is used, but without any scaling factors applied to the permittivity or ion mass, so as to correctly preserve the relevant spatial and temporal scales.

\subsection{Particle-In-Cell/Monte Carlo collisions (PIC/MCC) simulations}
\label{sec:2D_PIC}

LPPic2D uses the classical structure of a 2D PIC/MCC code. \cite{Birdsall} It features a structured Cartesian mesh, fixed in time, with square cells ($\Delta x = \Delta y$). The time-step, as well as the cell size, are  chosen so as to resolve the electron plasma frequency, the Debye length, and to satisfy the Courant-Friedrichs-Lewy (CFL) condition for electrons with a maximum energy of 150\,eV.

Ions and electrons are initialized with a uniform density, $n_0$, and with a given temperature ($T_e$ for electrons, and $T_i$ for ions). The number of particles initialized in the system, $N$, is a parameter chosen in order to obtain typically more than 60 particles-per-cell. \cite{Birdsall} The ions are assumed to be unmagnetized in HET due to their large Larmor radius compared with the dimensions of the system. Neutrals are not followed in the simulation but are treated as a constant and homogeneous background at given temperature $T_{n}$ and pressure $P_{n}$.

The region of a HET channel that is near the outlet of the thruster, where both the magnetic field and the acceleration electric field reach a maximum, is simulated. The curvature of the system is neglected, since it was shown that it does not play a significant role in the plasma discharge behavior. \cite{Trevor13} Therefore, the $(r-\theta)$ plane corresponds to the $(Oy-Ox)$ plane in Cartesian coordinates, with periodic boundary conditions in the $x$ direction. The $x$ and $y$ components of the electric field are obtained by solving Poisson's equation, while the $z$ component of the electric field is imposed as a parameter in the simulation. The system length in the $z$ direction is set to a finite value, $L_z$, but Poisson's equation is not solved in this direction. This method is a 2D generalization of previous 1D models \cite{TrevorI,Trevor38}, and has already been described in a former paper\cite{nous}. 

\subsection{Collisional processes}
\label{sec:collisions}

Collisional processes between charged particles and the neutral background are modeled using a Monte Carlo collision (MCC) algorithm \cite{Vahedi}. Electron-neutral collisions include elastic collisions, ionization, and several excitation processes. Since it would be very heavy to simulate the energy losses corresponding to all the excited states of the atom and that many cross section data remain unknown, sets of two to four inelastic processes that are consistent with the global behavior of the gas were selected from the literature. Ion-neutral elastic scattering and charge-exchange reactions are included for ions. All the electron-neutral collision cross sections are taken from the Biagi database retrieved from LXcat. \cite{biagi} Ion-neutral charge exchange cross sections for helium ($4.003$ AMU), argon ($39.95$ AMU) and xenon ($131.3$ AMU \cite{Xe-cr}) come from the Phelps database. \cite{phelps} For krypton ($83.8$ AMU), an empirical formula proposed by Sakabe is used. \cite{sakabe_simple_1992} The elastic collisions between ions and neutrals are modeled by Langevin capture cross sections,\cite{piscitelli, lieberman} except for argon for which data coming from Phelps \cite{phelps} is used. The set of cross section data corresponding to each gas is noted $\{\sigma_{X}\}$, where $X$ can be He, Ar, Kr, or Xe. This notation is used in particular in Table\,\ref{tab:cases}.

\subsection{Thruster walls}
\label{sec:HET-mod}

\begin{figure}[!tbp]
   \centering
   \includegraphics[width=0.9\linewidth]{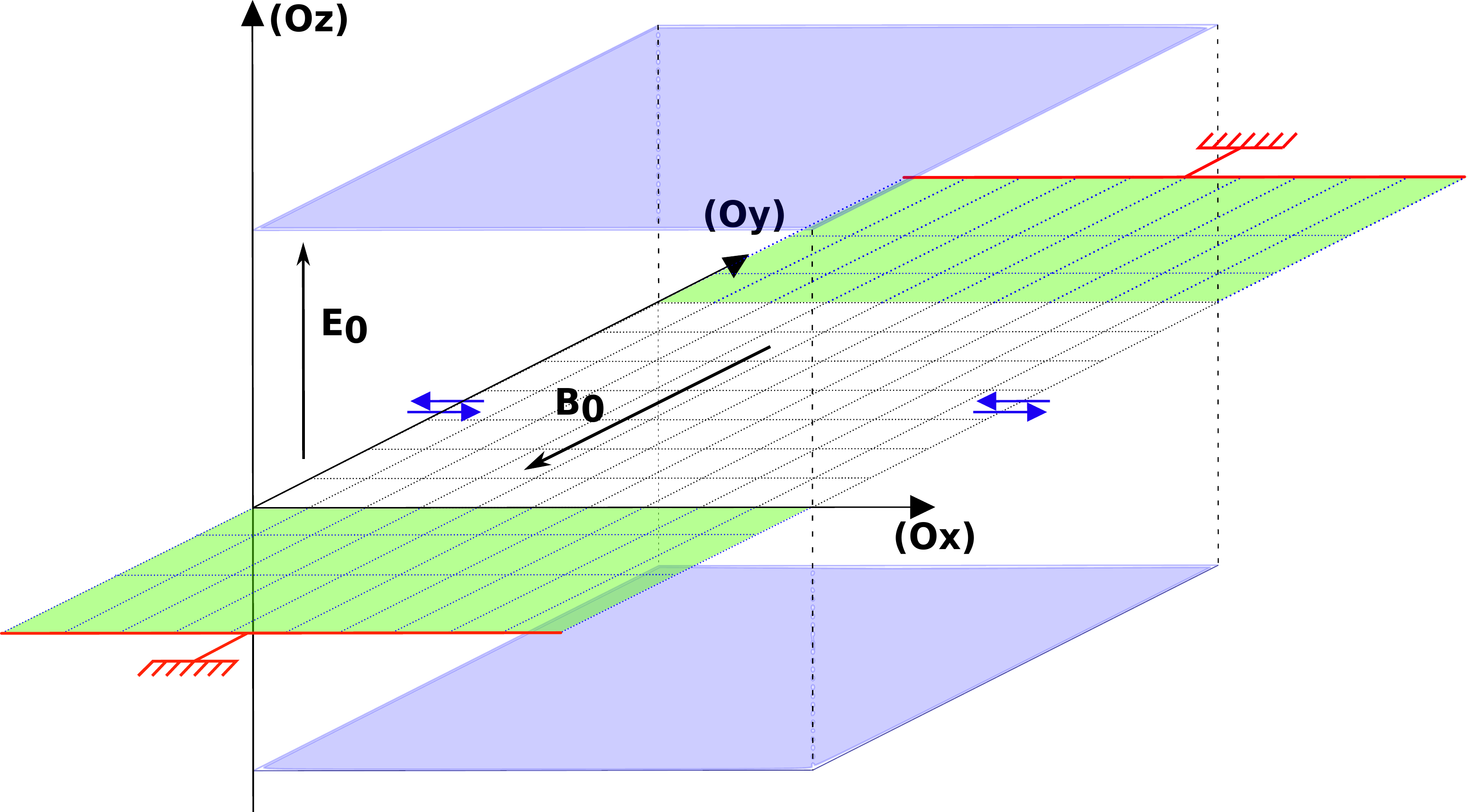}
   \caption{Schematic of the simulation set-up. The 2D $(Ox-Oy)$ grid is displayed with the plasma discharge in the center and the dielectric walls (purple) separating the discharge from grounded metallic electrodes (red). Blue arrows represent the periodic boundary conditions closing the $(Ox)$ axis. In green, the $z=0$ and $z=L_z$ planes close the simulation domain.}
   \label{fig:PIC-3D}
\end{figure}

Dielectric walls are modeled as a physical dielectric thickness on both sides of the channel that separate the discharge channel from a grounded metallic wall. The walls have the same characteristics (thickness, $L_{\mathrm{diel}}$, and relative permittivity, $\epsilon_{\mathrm{diel}}^r = \epsilon_{\mathrm{diel}}/\epsilon_0$) on both sides. These dielectrics are modeled by extending the simulation grid and solving Poisson's equation inside the dielectric layer while taking into account the surface charges that build up at the plasma-wall interface.\cite{Pechereau, pechereauthese,yue} This set-up is illustrated in Figure\,\ref{fig:PIC-3D}.

A linear model of SEE introduced by S. Barral \cite{SEE_Barral} and also used in more recent works \cite{Trevor13, CPhT_1, Syd-instability} is implemented. In this model, the incident electron kinetic energy, $\epsilon$, is used to estimate the re-emission probability:
\begin{equation}
   \label{eq:sigma_barral}
   \sigma(\epsilon) = \min( \, \sigma_0 + \frac{\epsilon}{\epsilon^{*}}[1-\sigma_0], \, \sigma_{\mathrm{max}} \, )
\end{equation}
where $\epsilon^{*}$ is the crossover energy, $\sigma_0$ is the minimum probability of secondary emission, and $\sigma_{\mathrm{max}}$ the maximum re-emission probability. Experimental studies of boron nitride walls, \cite{CPhT_1} which is the most commonly used material in HETs, show that typical values are $\sigma_0 = 0.578$, $\sigma_{\mathrm{max}} = 2.9$, and $\epsilon^* = 35.04$\,eV. 

\section{Results}
\label{sec:results}

The parameters used in the following simulations are summarized in Table \ref{tab:parameters}. In order to decouple the effect of the propellant from the one of the walls, two models are investigated: one simplified, {\it without} SEE and dielectric walls, and a more realistic case, {\it with} SEE and dielectric walls. 
\begin{table}
   \centering
   \begin{tabular}{@{}p{48mm}@{} c c}
   \\
      \toprule
      Parameter & Unit & Value \\
      \hline
      Gas & & Xe, Kr, Ar, He \\
      Simulation domain, $L_{x} \times L_{y} \times L_{z}$    & [cm$^3$] & $0.5 \times 2.0 \times 1.0$ \\
      Cell size, $\Delta x = \Delta y = \Delta z$  & [cm] & $2 \times 10^{-5}$ \\
      Time-step, $\Delta t$ & [s]  & $4 \times 10^{-12}$ \\
      Number of super-particles, $N$        & [particles]   & $25 \times 10^{6}$ \\
      Diagnostics average, $N_{A}$    & [time-steps]   & $2000$ \\
      Radial magnetic field, $B_{0}$    & [G]  & $200$ \\
      Axial electric field, $E_{0}$    & [$\textnormal{Vm}^{-1}$] & $2 \times 10^{4}$ \\
      Mean plasma density, $n_{0}$    & [$\textnormal{m}^{-3}$] & $3 \times 10^{17}$ \\
      Injection electron temperature, $T_{e}$    & [eV] & $5.0$ \\
      Injection ion temperature, $T_{i}$    & [eV] & $0.1$ \\
      SEE temperature, $T_{\mathrm{see}}$  & [eV] & $1.0$ \\
      Neutral gas pressure, $P_{n}$    & [mTorr]       & $1.0$ \\
      Neutral gas temperature, $T_{n}$    & [K]           & $300$ \\
      Neutral gas density, $n_{g}$    & [$\textnormal{m}^{-3}$] & $3.22 \times 10^{19}$ \\
      Probability of attachment, $\sigma_0$ &               & $0.578$ \\
      Maximum re-emission probability, $\sigma_{\mathrm{max}}$ &  & $2.9$ \\
      Crossover energy, $\epsilon^*$ & [eV]        & $35.04$ \\
      Dielectric thickness, $L_{\mathrm{diel}}$ & [mm]     & $3$ \\
      Dielectric relative permittivity, $\epsilon_{\mathrm{diel}}^r$ & & $10$ \\
      \botrule
   \end{tabular}
   \caption{ Physical and numerical parameters used in 2D PIC simulations.}
   \label{tab:parameters}
\end{table}
 
For each of these models, seven simulations are run. Four of them simulate the different propellants: xenon, argon, krypton, and helium (cases 1 and 2 in Table\,\ref{tab:cases}), and three other simulations are conducted with exactly the same set-up, but with collision processes of xenon instead of the corresponding gas collisions (case 3 in Table\,\ref{tab:cases}). In the simulations of case 3, the ions are simulated with their masses but with the xenon collision cross sections. These simulations are performed in order to isolate the effect of the ion mass. About 10\,$\mu$s of physical time are simulated, and the different values presented thereafter are averaged during the last $5 \,\mu$s of the simulation, when the instability has reached saturation.

\begin{table}[!tbp]
   \centering
\begin{tabular}{| l | c | c | c | c |}
\hline \multicolumn{1}{|c|}{\multirow{2}{*}{\bf Simulation case}} & \multicolumn{4}{c|}{\bf Ion mass [AMU]} \\ \cline{2-5}
      &   4.003           & 39.95             &    83.8           &    131.3          \\ \hline
{\bf Case 1}: &\multirow{4}{*}{$\{\sigma_{He}\}$} & \multirow{4} {*}{$\{\sigma_{Ar}\}$} & \multirow{4}{*}{$\{\sigma_{Kr}\}$} & \multirow{4}{*}{$\{\sigma_{Xe}\}$}\\ Dielectric walls with SEE \:   &  &&&  \\ \cline{1-1}
{\bf Case 2}: &&&&\\ 
Metallic walls without SEE &&&&      \\ \hline
{\bf Case 3}: & \multicolumn{4}{c|}{\multirow{2}{*}{$\{\sigma_{Xe}\}$}}  \\ 
Metallic walls without SEE &\multicolumn{4}{c|}{} \\ \hline
\end{tabular}
 \caption{The different simulation cases treated in terms of wall models and the implemented cross sections. $\{\sigma_{X}\}$ represents the set of cross section data corresponding to each gas, where $X$ can be He, Ar, Kr, or Xe.}
\label{tab:cases}
\end{table}

\subsection{Electron drift instability characteristics}

\begin{figure}[tbp]
\centering
\includegraphics[width = 0.9\linewidth]{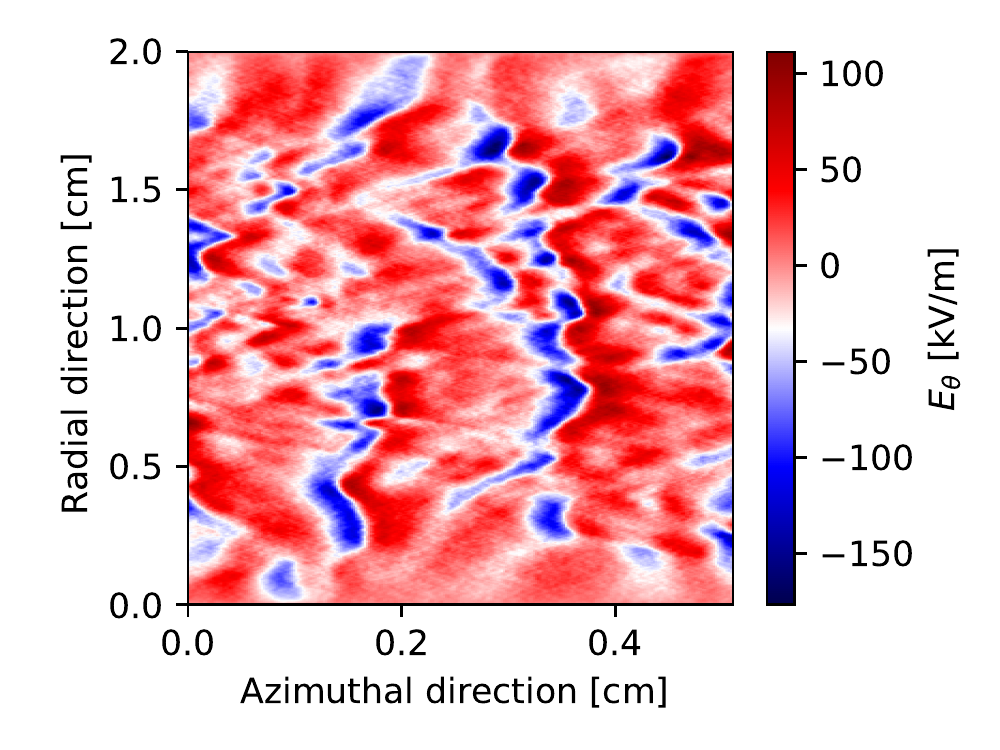}
\caption{Snapshot at $10~\mu s$ of the azimuthal electric field $ E_{\theta} = \mathbf{E} \cdot \mathbf{e_{\theta}}$ in kV/m in the 2D $(r-\theta)$ domain with the simplified model and xenon as propellant.}
\label{fig:Ex_Snapshot}
\end{figure}

\begin{figure}[tbp]
\centering
\includegraphics[width = 0.9\linewidth]{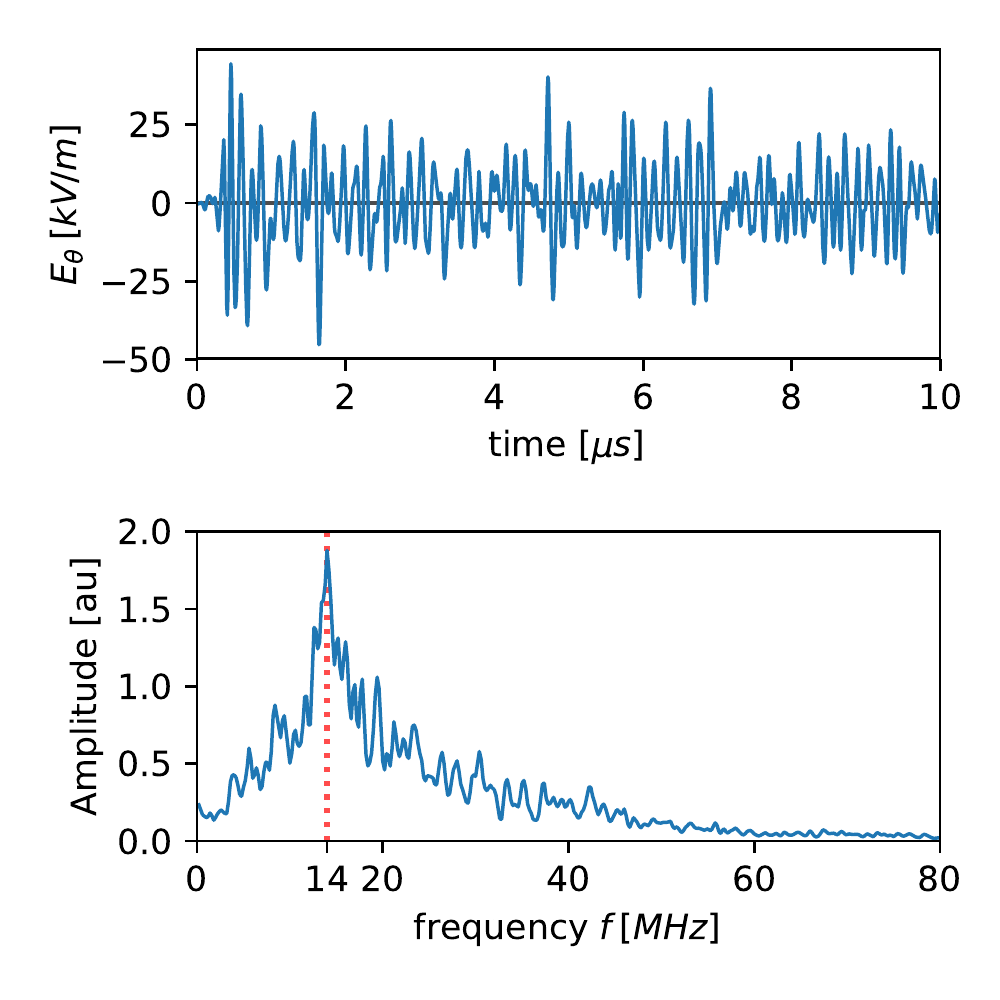}
\caption{ (top) Temporal evolution of the azimuthal electric field $ E_{\theta}$ at the center of the simulation domain for argon (case 1) and (bottom) the corresponding frequency spectrum. The main frequency is clearly observed, here at 14\,MHz.
\label{fig:FFT}
}
\end{figure}

In each case, an instability grows in less than 2\,$\mu$s (the growth time depending of the gas) before it saturates. The instability qualitatively has the same characteristics as those found in previous studies using xenon. \cite{Trevor10, Trevor11, Trevor12, Trevor13} An example of the instability is illustrated with the azimuthal electric field $E_{\theta}$ in Figure\,\ref{fig:Ex_Snapshot} and Figure\,\ref{fig:FFT}. Figure\,\ref{fig:Ex_Snapshot} illustrates the instability spatial pattern, with the main azimuthal oscillation and smaller oscillations in both the radial and the azimuthal directions. Figure\,\ref{fig:FFT} present the temporal evolution of $E_{\theta}$ and the corresponding frequency spectrum. A more detailed description of the simulation case with xenon can be found in a previous paper. \cite{croes_modelisation_2017, nous}

\begin{figure}[!tbp]
	\centering
	\includegraphics[width = 0.85\linewidth]{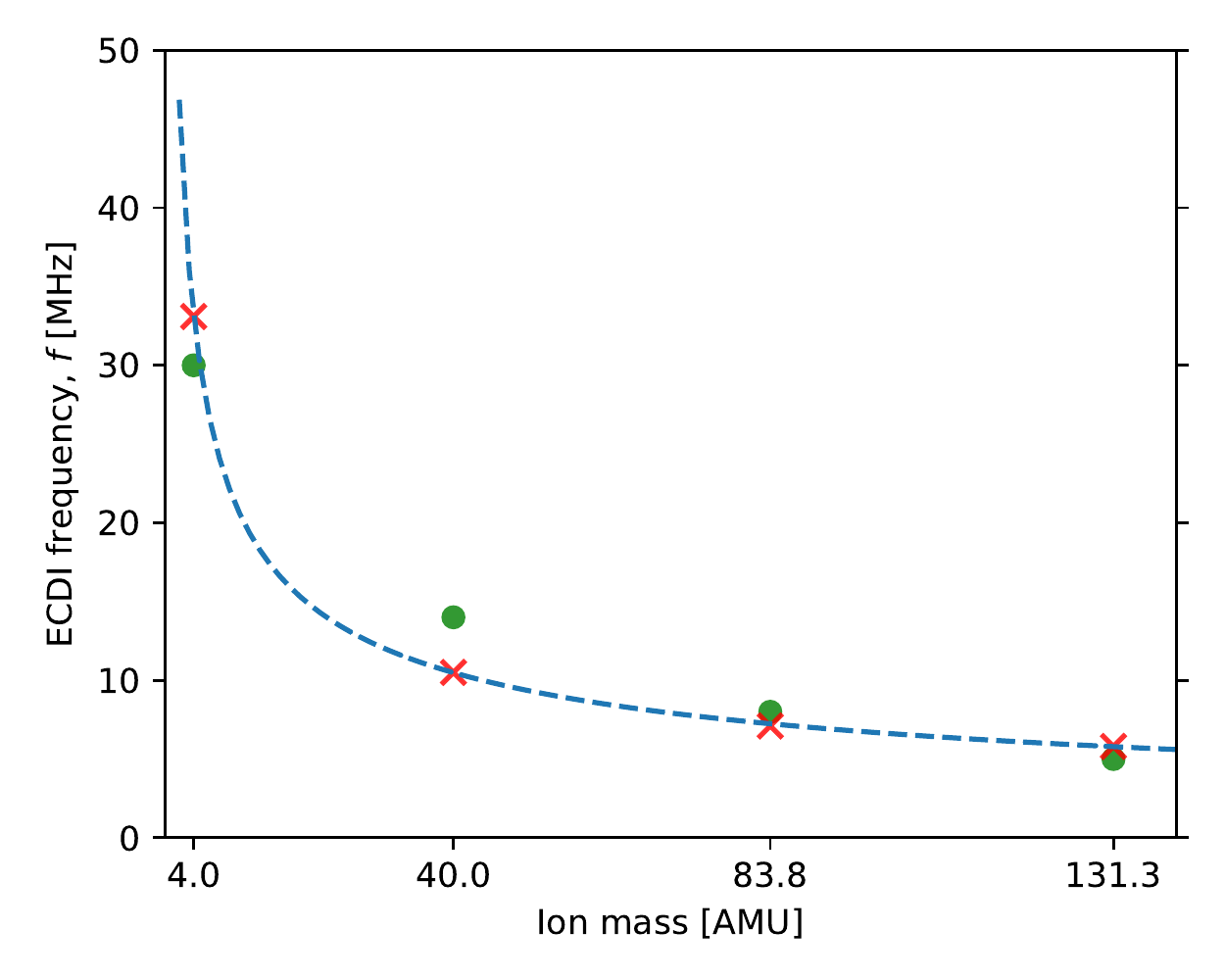} \\
	\caption{Comparison between (dashed blue line) the theoretical value of the electron drift instability frequency $f$ obtained from equation (\ref{eq:f}), and (markers) the values obtained from the PIC/MCC simulations: $\times$: case 1, $\bullet$: cases 2 \& 3 (overlapping).}
    \label{fig:f-comp} 
\end{figure}

Instability properties such as the frequency, $f$, or the wavelength, $\lambda$, are defined by the locations of the maxima in the temporal and spatial Fourier spectra, respectively. The azimuthal Fourier transform of the data is averaged in the radial direction (excluding the sheaths). The maximum frequency and wavelength are computed for each run and compared to the theoretical predictions from kinetic theory derived in Lafleur {\it et al.}\cite{TrevorII}, where it was shown that:
\begin{align}
   f & =  \frac{\omega_{pi}}{2 \pi\sqrt{3}} \qquad & \text{with} & \qquad  \omega_{pi}  = \left( \frac{n_0 q^2}{\epsilon_{0} m_{i}} \right)^{1/2}  \label{eq:f} \\
   \lambda & =  2 \pi \lambda_{De}\sqrt{2} \qquad & \text{with} & \qquad \lambda_{De} = \left( \frac{\epsilon_{0} T_{e}}{|q|n_0} \right)^{1/2} \label{eq:lambda}
\end{align}
where $\lambda_{De}$ is the electron Debye length and $\omega_{pi}$ the ion plasma frequency.
The uncertainty in the measurements from the PIC simulation data are approximately $\pm 0.5$\,MHz for the frequency and $\pm 0.1$\,mm for the wavelength. Moreover, the measurement of the electron temperature, calculated directly from the mean electron kinetic energy, has an uncertainty of approximately $\pm 5$\,eV. This uncertainty is then echoed in any subsequent estimates, such as the Debye length. Frequency comparisons are summarized in Figure \ref{fig:f-comp}. The instability frequency decreases with the ion mass in agreement with equation\,(\ref{eq:f}) displayed as the dashed line in Figure\,\ref{fig:f-comp}. The wavelength of the instability is observed to be around $1.1$\,mm for all cases, which is close to the value  predicted by equation (\ref{eq:lambda}).  

As already demonstrated for a xenon plasma discharge, \cite{nous} collision processes  play only a minor role in the characteristics of the instability. Indeed, comparison between simulation cases 2 and 3 show very weak discrepancies. This agreement confirms that ion-neutral and electron-neutral collisions represent only higher order effects on the instability.

\subsection{Enhanced electron transport}

In the PIC simulations, the mean cross field electron mobility is measured in the direction perpendicular to the simulation plane according to the formula:
\begin{equation}
   \label{eq:mu-pic}
   \mu_{\mathrm{pic}}=\frac{\sum_{j=1}^{\mathscr{N}_{e^{-}}}v_{jz}}{\mathscr{N}_{e^{-}}E_{0}}
\end{equation}
where the summation is performed over all the $\mathscr{N}_{e^{-}}$ electrons, and $v_{jz}$ is the $z$ component of the velocity of the $j^{th}$ electron. According to classical transport theory, the electron mobility should be:\cite{lieberman}
\begin{equation}
   \label{eq:classical-mu}
   \mu_{\mathrm{cla}} = \frac{\frac{e}{m_{e} \nu_{m}}}{1+\frac{\omega_{ce}^{2}}{\nu_{m}^{2}}}
\end{equation}
where $\omega_{ce}=eB_{0}/m_{e}$ is the electron cyclotron frequency, and $\nu_{m}$ is the electron-neutral momentum transfer collision frequency which is measured directly from the PIC simulation.
An approximate improved cross-field electron mobility found by Lafleur \cite{TrevorI} is
\begin{equation}
   \label{eq:mu-eff}
   \mu_{\mathrm{eff}} = \mu_{\mathrm{cla}} \left( 1- \frac{\omega_{ce}}{\nu_{m}} \frac{\langle n_{e} E_{\theta} \rangle}{n_0 E_{0}} \right)
\end{equation}
where $n_{e}$ is the local electron density, $E_{\theta}$ the electric field in the azimuthal direction, and $\langle \cdot \rangle$ represents a spatial and temporal average. Under the assumption that the saturation of the instability is mainly due to ion trapping, the electron mobility may be simplified to\cite{TrevorII}:
\begin{equation}
\label{eq:mu-sat1}
\mu_{\mathrm{eff}}^{sat} = \mu_{\mathrm{cla}}\left( 1 + \frac{\omega_{ce}}{\nu_m}\frac{R_{ei}}{|q|n_eE_0}  \right)
\end{equation}
where
\begin{equation}
   \label{eq:Rei}
   R_{ei} = \frac{e|\nabla\cdot(n_e T_e \mathbf{v_i})|}{4\sqrt{6} c_s} 
   \approx \frac{e n_e T_e v_{iz}}{4\sqrt{6} c_s L_z}
\end{equation}
is the saturated electron-ion friction force. In equation (\ref{eq:Rei}), the spatial derivative has been approximated across the axial simulation direction, $\mathbf{v}_i$ is the local ion drift velocity, and $c_s = (e T_e/m_i)^{1/2}$ is the Bohm speed. The ion outlet velocity along $(Oz)$ in the simulation is:
\begin{equation}
  \label{eq:vzi}
  v_{iz} = \left(\frac{2 e E_{0} L_{z}}{m_i}\right)^{1/2}
\end{equation}
In a real thruster, the ion velocity is driven by the electrostatic potential drop between the anode and the cathode, $U$ ($U>0$). In the simulation, since Poisson's equation is not solved in the axial direction, the potential drop is related to the applied axial electric field from
\begin{equation}
E_0 L_z = U
\end{equation}

Equation (\ref{eq:mu-sat1}) hence becomes:
\begin{equation}
   \label{eq:mu-sat2}
   \mu_{\mathrm{eff}}^{sat} = \mu_{\mathrm{cla}}\left[ 1 + \frac{1}{4\sqrt{3}}\frac{\omega_{ce}}{\nu_m}\left( \frac{T_e}{U} \right)^{1/2} \right] 
\end{equation}
Since $\nu_m \ll \omega_{ce}$, equation (\ref{eq:mu-sat2}) reduces to:
\begin{equation}
\label{eq:mu-sat3}
 \mu_{\mathrm{eff}}^{sat} = \frac{(T_e/U)^{1/2}}{4\sqrt{3}B_0}
\end{equation}
Equation (\ref{eq:mu-sat3}) shows that to this approximation the enhanced mobility does not explicitly depend on the considered gas.
Interestingly, with the approximation to the derivative in equation\,(\ref{eq:Rei}), the enhanced mobility shows a $1/B$ dependence; similar to that for a Bohm-type anomalous transport. This however is somewhat fortuitous as the applied magnetic field is spatially constant, and the electron mobility has been averaged over the axial simulation domain length. For the case where the spatial variation of the magnetic field and electron mobility are accounted for, this $1/B$ dependence is no longer satisfied in general. \cite{TrevorII, Trevor5}

\begin{figure}[!tbp]
	\centering
	\includegraphics[width = 0.95\linewidth]{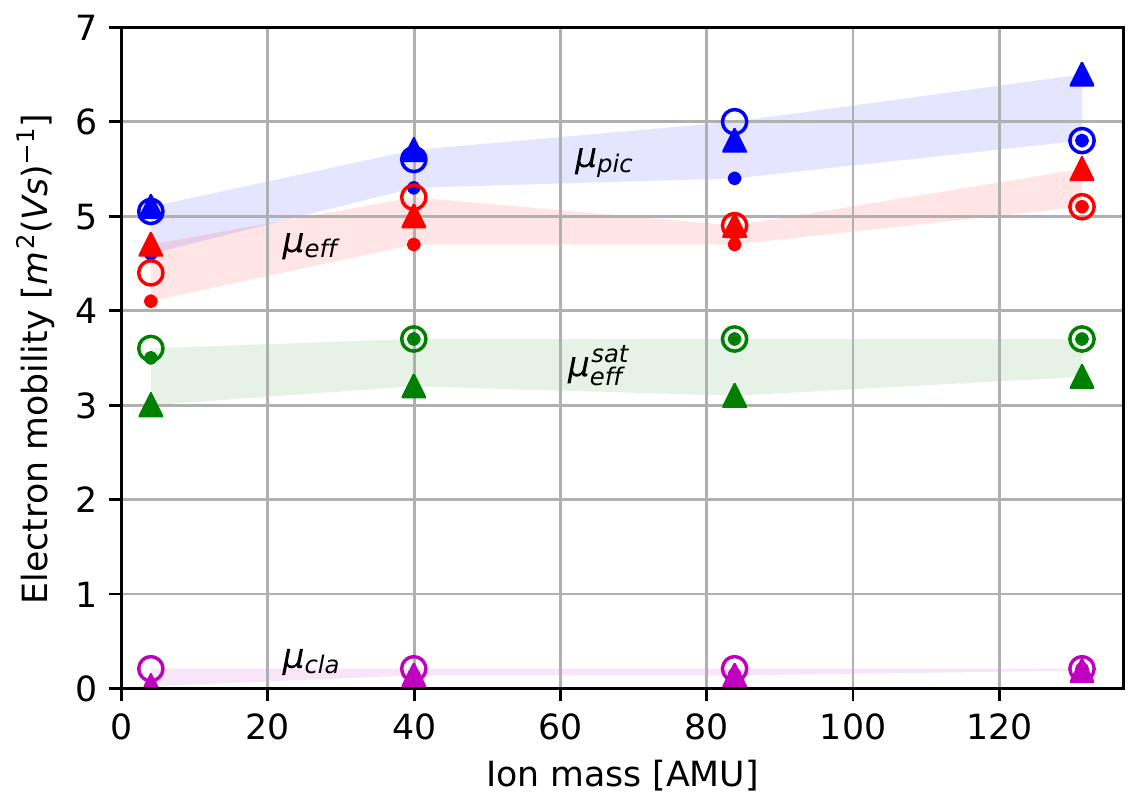}
	\caption{Evolution of the mobility estimates and results from the PIC simulation: (blue) $\mu_{\mathrm{pic}}$ given by equation (\ref{eq:mu-pic}), (red) $\mu_{\mathrm{eff}}$ given by equation (\ref{eq:mu-eff}), (green) $\mu_{\mathrm{eff}}^{sat}$ given by equation (\ref{eq:mu-sat2}), and (purple) $\mu_{\mathrm{cla}}$ given by equation (\ref{eq:classical-mu}).  $\blacktriangle$: case 1, $\bigcirc $: case 2, $\bullet$: case 3 }
	\label{fig:mob-comp}
\end{figure}

A comparison between the measured and theoretical values is shown in Figure\,\ref{fig:mob-comp}. The enhanced mobility measured from the PIC simulation slightly increases with the mass of the propellant. It varies from 5\,$\textnormal{m}^{2} \textnormal{V}^{-1} \textnormal{s}^{-1}$ for helium to 6\,$\textnormal{m}^{2} \textnormal{V}^{-1} \textnormal{s}^{-1}$ for xenon. Error margins on the computation of the mobilities are estimated to be $\pm 0.1 \, \textnormal{m}^{2} \textnormal{V}^{-1} \textnormal{s}^{-1}$. The effective mobility predicted by equation (\ref{eq:mu-eff}) agrees well with the PIC simulation results. The saturated effective mobility (equations (\ref{eq:mu-sat2}-\ref{eq:mu-sat3})) shows a reasonable agreement with the mobility measured from the PIC simulation, and provides some scaling laws of the mobility with respect to the discharge voltage, the electron temperature, and the magnetic field.  

As expected from kinetic theory, even though the frequency decreases with the ion mass, the enhanced mobility measured in the simulation is not heavily impacted by a change of propellant. However, in a real system where the electron temperature is determined by particle and energy balance considerations inside the discharge, the electron temperature may vary as the propellant mass changes.

\section{Summary and conclusion}
\label{sec:discuss}

\subsection{Role of the propellant}
In the previous sections, the results of 2D PIC/MCC simulations in the $(r-\theta)$ plane of a HET operated with various gases were presented. Enhanced electron transport as well as the electron drift instability have been observed and quantified. 
These simulations used various noble gases (helium, argon, krypton, and xenon) in order to investigate the impact of the propellant on the plasma discharge.
It was verified that the influence of the propellant on the characteristics of the EDI are well predicted by kinetic theory. A simplified formula for the anomalous electron mobility (equation (\ref{eq:mu-sat3})) was proposed and matches reasonably well with the PIC simulation results. However, this theoretical expression is not able to predict the slight increase of the electron mobility with the propellant mass.

\subsection{Role of the wall properties}

The first impact of SEE is to cool down the plasma. Indeed, secondary electrons are injected in the sheath, reducing its potential drop. Hence the minimum electron energy needed to exit the system is lower. Moreover, secondary electrons are injected at a temperature of $T_{\mathrm{see}}=1$\,eV which is much lower than the temperature of the bulk electrons.
For example, for argon, the electron mean temperature in the bulk drops from 48.3\,eV to 36.2\,eV when the SEE are taken into account. 
Several runs were also performed with dielectric walls and without the SEE in order to discriminate between the effect of the SEE and the dielectric layer.
It was found that the electron temperature is not affected whether the walls feature a dielectric coating or not. According to our theory (equation\,\ref{eq:mu-sat3}), the mobility increases with the electron temperature, so a lower mobility is predicted when the SEEs are taken into account ($\mu_{\text{eff}}^{sat}$ in Figure\,\ref{fig:mob-comp}).
However, a space-resolved analysis of the electron mobility in the $y$ direction shows that the mobility is not uniform, so that the averaged model derived from equation (\ref{eq:mu-sat1}) is not valid anymore.
The mobility is actually about twice higher in the sheath region than in the bulk. This result was already found in previous studies and interpreted as a near-wall mobility effect. \cite{Pustovitov2000} This phenomenon is caused by the SEE and its effect on the averaged mobility is stronger than that of the drop of electron temperature. 
Consequently, the values of electron mobility measured from the PIC simulation data and predicted by equation\,(\ref{eq:mu-eff}) are higher in case\,1 than in cases\,2 and 3.


\subsection{Future work}
The main limitation of this work is that Poisson's equation is only solved in the $(r-\theta)$ plane. This implies that the $z$ component of the wavenumber is neglected and that the convection of the instability away from the simulation plane is not correctly modeled. Moreover, self-consistent ionization was not possible in our configuration as the ionization zone of an HET differs from the acceleration zone. Therefore, the effect of the changed chemistry on the ionization process should be investigated with the axial direction taken into account.

Moreover, as the present simulation tool does not allow us to model more than one ion type, we are unable to model complex chemistry processes. Thus, atmospheric propellants as well as iodine are promising alternatives which will be studied in a future work.

Finally, through comparisons of the PIC/MCC simulation results with a kinetic theory coming from recent literature,\cite{TrevorI,TrevorII,nous} we showed that the electron drift instability is the main driver of the plasma transport in HET operated with noble gas propellants. We provided useful estimates and analytical formulas that can be implemented in future global models of HET that would capture the main trends of the PIC simulation. 

\section*{Acknowledgements}

The authors would like to thank Francois Pechereau, Abdoul Wahid C. Mainassara, and Stephan Zurbach for numerous useful discussions, as well as for their help and support in the development of the LPPic2D simulation code. V.C. and A.T. acknowledge financial and technical support from a Safran Aircraft Engines (formerly SNECMA) doctoral research award. This work has been partially funded by the CHEOPS project that has received funding from the European
Union’s Horizon 2020 research and innovation program under grant agreement No 730135, as well as from the Association Nationale de la Recherche et de la Technologie (ANRT). We were granted access to the HPC resources of CINES under the allocation 2016-AP010510091 made by GENCI.

\section*{References}

\bibliographystyle{unsrt}
\bibliography{biblio}

\begin{thebibliography}{10}

\bibitem{business}
R.~Villain.
\newblock {\em {Satellites to be built and launched by 2014}}, volume~18.
\newblock Euroconsult Research Report, 2015.

\bibitem{Goebel}
D.~M. Goebel and I.~Katz.
\newblock {\em {Fundamentals of Electric Propulsion: Ion and Hall Thrusters}}.
\newblock Wiley, 2008.

\bibitem{USpatent}
D.Q. King, K.H. de~Grys, R.S. Aadland, D.L. Tilley, and A.W. Voigt.
\newblock {Magnetic flux shaping in ion accelerators with closed electron
  drift}, March~27 2001.
\newblock US Patent 6,208,080.

\bibitem{geometry}
A.~Bouchoule, A.~Cadiou, A.~Heron, M.~Dudeck, and M.~Lyszyk.
\newblock {An overview of the French research program on plasma thrusters for
  space applications}.
\newblock {\em Contrib. Plasma Phys.}, 41(573), 2001.

\bibitem{Trevor5}
J.~C. Adam, J.~P. Boeuf, N.~Dubuit, M.~Dudeck, L.~Garrigues, D.~Gresillon,
  A.~Heron, G.~J.~M. Hagelaar, V.~Kulaev, N.~Lemoine, S.~Mazouffre,
  J.~Perez-Luna, V.~Pisarev, and S.~Tsikata.
\newblock {Physics, simulation, and diagnostics of Hall effect thrusters}.
\newblock {\em Plasma Phys. Control. Fusion}, 24(124041), 2008.

\bibitem{nous}
V.~Croes, T.~Lafleur, Z.~Bonaventura, A.~Bourdon, and P.~Chabert.
\newblock {2D particle-in-cell simulations of the electron drift instability
  and associated anomalous electron transport in Hall-effect thrusters}.
\newblock {\em Plasma Sources Sci. Technol.}, 26(034001), 2017.

\bibitem{Mazouffre}
S.~Mazouffre.
\newblock Electric propulsion for satellites and spacecraft: established
  technologies and novel approaches.
\newblock {\em Plasma Sources Science and Technology}, 25(3):033002, 2016.

\bibitem{Anne01}
E.~Ahedo.
\newblock {Plasmas for space propulsion}.
\newblock {\em Plasma Phys. Control. Fusion}, 53(124037), 2011.

\bibitem{2stage2}
R.~R. Hofer, P.~Y. Peterson, and A.~D. Gallimore.
\newblock {A High Specific Impulse Two-Stage Hall Thruster with Plasma Lens
  Focusing}.
\newblock In {\em {IEPC-01-036}}, 2001.

\bibitem{2stages}
C.~Boniface, G.~J.~M. Hagelaar, L.~Garrigues, J.~P. Boeuf, and M.~Prioul.
\newblock {Modeling of double stage Hall effect thruster}.
\newblock {\em IEEE Transactions on Plasma Science}, 33(2):522--523, April
  2005.

\bibitem{Krypton}
A.~Kieckhafer and L.~B. King.
\newblock {Energetics of propellant options for high-power Hall thrusters}.
\newblock {\em J. Propul. Power}, 23:21--26, 2007.

\bibitem{I2-prop}
R.~Dressler, Y.~H. Chiu, and D.~Levandier.
\newblock {Propellant alternatives for ion and Hall effect thrusters}.
\newblock In {\em {AIAA-2000-0602}}, {38th Aerospace Sciences Meetings}.
  American Institute of Aeronautics and Astronautics, jan 2000.

\bibitem{I2-exp}
J.~Szabo, B.~Pote, S.~Paintal, M.~Robin, A.~Hillier, D.~R. Branam, and R.~E.
  Huffmann.
\newblock {Performance Evaluation of an Iodine-Vapor Hall Thruster}.
\newblock {\em J. Propul. and Power}, 28(4):848--857, 2012.

\bibitem{I2}
P.~Grondein, T.~Lafleur, P.~Chabert, and A.~Aanesland.
\newblock {Global model of an iodine gridded plasma thruster}.
\newblock {\em Phys. Plasmas}, 23(033514), 2016.

\bibitem{air}
G.~Cifali, D.~Dignani, T.~Misuri, P.~Rossetti, M.~Andrenucci, D.~Valentian,
  F.~Marchandise, D.~Feili, and B.~Lotz.
\newblock {Completion of HET and RIT characterization with atmospheric
  propellants}.
\newblock In {\em Proceedings of Space Propulsion, Bordeaux}, 2012.

\bibitem{ESA1}
S.~Barral and L.~Walpot.
\newblock {Conceptual of an Air-Breathing Electric Propulsion System}.
\newblock In {\em {IEPC-2015-271}}, 2015.

\bibitem{Trevor1}
A.~I. Morozov, V.~V. Savelyev edited by B.~B. Kadomtsev, and V.~D. Shafranov.
\newblock {\em {Reviews of Plasma Physics}}.
\newblock Springer Science+Business Media, New-York, 2000.

\bibitem{Trevor4}
N.~B. Meezan, W.~A.~Jr Hargus, and M.~A. Cappelli.
\newblock {Anomalous electron mobility in a coaxial Hall discharge plasma}.
\newblock {\em Phys. Rev., E 63, 026410}, 63(2), 2001.

\bibitem{HET-transport-measures}
N.~B. Meezan.
\newblock {\em {Electron transport in a coaxial Hall discharge}}.
\newblock Ph.d., {Department of Mechanical Engineering, Stanford University},
  2002.

\bibitem{Trevor6}
L.~Garrigues, J.~P{\'e}rez-Luna, J.~Lo, G.~J.~M. Hagelaar, J.~P. Boeuf, and
  S.~Mazouffre.
\newblock {Empirical electron cross-field mobility in a Hall effect thruster}.
\newblock {\em Appl. Phys. Lett.}, 95(141501), 2009.

\bibitem{Trevor7}
I.~Katz, I.~G. Mikellides, B.~A. Jorns, and A.~L. Ortega.
\newblock {Hall2De simulations with an anomalous transport model based on the
  electron cyclotron drift instability}.
\newblock In {\em {IEPC-2015-402}}, 2015.

\bibitem{Trevor8}
I.~D. Kaganovich, Y.~Raitses, D.~Sydorenko, and A.~Smolyakov.
\newblock {Kinetic effects in a Hall thruster discharge}.
\newblock {\em Phys. Plasmas}, 14(057104), 2007.

\bibitem{Trevor9}
D.~Sydorenko, A.~Smolyakov, I.~Kaganovitch, and Y.~Raitses.
\newblock {Electron kinetic effects and beam related instabilities in Hall
  thrusters}.
\newblock {\em Phys. Plasmas}, 15(053506), 2008.

\bibitem{Trevor18}
N.~Gascon, M.~Dudeck, and S.~Barral.
\newblock {Wall material effects in stationary plasma thrusters. I. Parametric
  studies of an SPT-100}.
\newblock {\em Phys. Plasmas}, 10(4123), 2003.

\bibitem{Trevor10}
M.~Hirakawa.
\newblock {Electron transport mechanism in a Hall thruster}.
\newblock In {\em {IEPC-97-021}}, 1997.

\bibitem{Trevor19}
C.~Boniface, L.~Garrigues, G.~J.~M. Hagelaar, J.~P. Boeuf, D.~Gawron, and
  S.~Mazouffre.
\newblock {Anomalous cross-field electron transport in a Hall thruster}.
\newblock {\em Appl. Phys. Lett.}, 89(161503), 2006.

\bibitem{Trevor20}
G.~J.~M. Hagelaar, J.~Bareilles, L.~Garrigues, and J.~P. Boeuf.
\newblock {Role of anomalous electron transport in a stationary plasma thruster
  simulation}.
\newblock {\em J. Appl. Phys.}, 93(67), 2003.

\bibitem{Trevor21}
N.~B. Meezan and M.~A. Capelli.
\newblock {Kinetic Study of Wall Collisions in a Coaxial Hall Discharge}.
\newblock {\em Phys. Rev.}, E 66(036401), 2002.

\bibitem{Trevor22}
F.~I. Parra, E.~Ahedo, J.~M. Fife, and M.~Martinez-Sanchez.
\newblock {A two-dimensional hybrid model of the hall thruster discharge}.
\newblock {\em J. Appl. Phys.}, 100(023304), 2006.

\bibitem{Trevor23}
A.~N. Smirnov, Y.~Raitses, and N.~J. Fisch.
\newblock {Electron cross-field transport in a miniaturized cylindrical Hall
  thruster}.
\newblock {\em IEEE Trans. Plasma Sci.}, 34(132), 2006.

\bibitem{Trevor24}
S.~Yoshikawa and D.~J. Rose.
\newblock {Anomalous Diffusion of a Plasma across a Magnetic Field}.
\newblock {\em Phys. Fluids}, 5(334), 1962.

\bibitem{Trevor26}
E.~Y. Choueiri.
\newblock {Fundamental difference between the two Hall thruster variants}.
\newblock {\em Phys. Plasmas}, 8(1411):5025--5033, 2001.

\bibitem{Trevor27}
A.~W. Smith and M.~A. Cappelli.
\newblock {Time and space-correlated plasma potential measurements in the near
  field of a coaxial Hall plasma discharge}.
\newblock {\em Phys. Plasmas}, 16(073504), 2009.

\bibitem{Trevor28}
M.~K. Scharfe, N.~Gascon, M.~A. Cappelli, and E.~Fernandez.
\newblock {Comparison of hybrid Hall thruster model to experimental
  measurements}.
\newblock {\em Phys. Plasmas}, 13(083505), 2006.

\bibitem{Trevor11}
J.~C. Adam, A.~H{\'e}ron, and G.~Laval.
\newblock {Study of stationary plasma thrusters using two-dimensional fully
  kinetic simulations}.
\newblock {\em Phys. Plasmas}, 11(295), 2004.

\bibitem{Trevor12}
A.~Ducrocq, J.~C. Adam, A.~H{\'e}ron, and G.~Laval.
\newblock {High-frequency electron drift instability in the cross-field
  configuration of Hall thrusters}.
\newblock {\em Phys. Plasmas}, 13(102111), 2006.

\bibitem{Trevor29}
J.~Cavalier, N.~Lemoine, G.~Bonhomme, S.~Tsikata, C.~Honore, and D.~Gresillon.
\newblock {Hall thruster plasma fluctuations identified as the {E x B} electron
  drift instability : Modeling and fitting on experimental data.}
\newblock {\em Phys. Plasmas}, 20(082107), 2013.

\bibitem{TrevorI}
T.~Lafleur, S.~D. Baalrud, and P.~Chabert.
\newblock {Theory for the anomalous electron transport in Hall effect
  thrusters: I Insights from particle-in-cell simulations}.
\newblock {\em Phys. Plasmas}, 23(053502), 2016.

\bibitem{TrevorII}
T.~Lafleur, S.~D. Baalrud, and P.~Chabert.
\newblock {Theory for the anomalous electron transport in Hall effect
  thrusters: II. Kinetic model}.
\newblock {\em Phys. Plasmas}, 23(053503), 2016.

\bibitem{Trevor37}
P.~Coche and L.~Garrigues.
\newblock {A two-dimensional (azimuthal-axial) Particle-In-Cell model of a Hall
  thruster}.
\newblock {\em Phys. Plasmas}, 21(023503), 2014.

\bibitem{Xenon}
R.~Betzendahl.
\newblock {The 2014 rare gases market report}.
\newblock {\em Cryogas International}, 52(28), 2014.

\bibitem{croes_modelisation_2017}
V.~Croes.
\newblock {\em Mod\'elisation bidimensionnelle de la d\'echarge plasma dans un
  propulseur de {Hall}}.
\newblock PhD thesis, Universit\'e Paris-Saclay, Palaiseau, 2017.

\bibitem{croes_study_2017}
V.~Croes, A.~Tavant, R.~Lucken, T.~Lafleur, A.~Bourdon, and P.~Chabert.
\newblock Stduy of electron transport in a {Hall} effect thruster with {2D}
  {$r-\theta$} {Particle}-{In}-{Cell} simulations.
\newblock In {\em IEPC-2017-57}, 2017.

\bibitem{lucken_edge--center_2018}
R.~Lucken, V.~Croes, T.~Lafleur, J.-L. Raimbault, A.~Bourdon, and P.~Chabert.
\newblock Edge-to-center plasma density ratios in two-dimensional plasma
  discharges.
\newblock {\em Plasma Sources Sci. Technol.}, 27(3), 2018.

\bibitem{lucken_global_2017}
R.~Lucken, V.~Croes, T.~Lafleur, J.-L. Raimbault, A.~Bourdon, and P.~Chabert.
\newblock Global models of plasma thrusters: {Insights} from {PIC} simulation
  and fluid theory.
\newblock In {\em IEPC-2017-323}, 2017.

\bibitem{Birdsall}
C.~K. Birdsall and A.~B. Langdon.
\newblock {\em {Plasma Physics via Computer Simulation}}.
\newblock McGraw-Hill, New-York, 1985.

\bibitem{Trevor13}
A.~H{\'e}ron and J.~C. Adam.
\newblock {Anomalous conductivity in Hall thrusters : Effects of the non-linear
  coupling of the electron-cyclotron drift instability with secondary electron
  emission of the walls.}
\newblock {\em Phys. Plasmas}, 20(082313), 2013.

\bibitem{Trevor38}
J.~P. Boeuf.
\newblock {Rotating structures in low temperature magnetized plasmas---insight
  from particle simulations}.
\newblock {\em Front. Phys.}, 2(74), 2014.

\bibitem{Vahedi}
V.~Vahedi and M.~Surendra.
\newblock {A {M}onte {C}arlo collision model for the particle-in-cell method:
  applications to argon and oxygen discharges}.
\newblock {\em Comp. Phys. Commun.}, 87(179), 1995.

\bibitem{biagi}
S.~F. Biagi.
\newblock Programm magboltz v7.1. cross section compilation. see www.lxcat.net.
\newblock retrieved on November 16, 2016, 2004.

\bibitem{Xe-cr}
A.~V. Phelps.
\newblock {Compilation of atomic and molecular data}.
\newblock 2005.

\bibitem{phelps}
A.~V. Phelps.
\newblock The application of scattering cross sections to ion flux models in
  discharge sheaths.
\newblock {\em Journal of Applied Physics}, 76(2):747, June 1998.

\bibitem{sakabe_simple_1992}
S.~Sakabe and Y.~Izawa.
\newblock Simple formula for the cross sections of resonant charge transfer
  between atoms and their positive ions at low impact velocity.
\newblock {\em Physical Review A}, 45(3):2086, 1992.

\bibitem{piscitelli}
D.~Piscitelli, A.~V. Phelps, J.~de~Urquijo, E.~Basurto, and L.~C. Pitchford.
\newblock Ion mobilities in {Xe}/{Ne} and other rare-gas mixtures.
\newblock {\em Physical Review E}, 68(4), 2003.

\bibitem{lieberman}
M.A. Lieberman and A.J. Lichtenberg.
\newblock {\em {Principles of plasma discharges and materials processing}}.
\newblock Wiley, 1994.

\bibitem{Pechereau}
F.~Pechereau, Z.~Bonaventura, and A.~Bourdon.
\newblock Influence of surface emission processes on a fast-pulsed dielectric
  barrier discharge in air at atmospheric pressure.
\newblock {\em Plasma Sources Science and Technology}, 25(4):044004, 2016.

\bibitem{pechereauthese}
F.~Pechereau.
\newblock {\em {Numerical simulation of the interaction of atmospheric pressure
  plasma discharges with dielectric surfaces}}.
\newblock Theses, {Ecole Centrale Paris}, December 2013.

\bibitem{yue}
Y.~Liu, J.-P. Booth, and P.~Chabert.
\newblock Plasma non-uniformity in a symmetric radiofrequency
  capacitively-coupled reactor with dielectric side-wall: a two dimensional
  particle-in-cell/monte carlo collision simulation.
\newblock {\em Plasma Sources Science and Technology}, 27(2):025006, 2018.

\bibitem{SEE_Barral}
S.~Barral, K.~Makowski, Z.~Peradznski, N.~Gascon, and M.~Dudeck.
\newblock {Wall material effects in stationary plasma thrusters. II. Near-wall
  and in-wall conductivity}.
\newblock {\em Phys. Plasmas}, 10(4137), 2003.

\bibitem{CPhT_1}
A.~N. Smirnov, Y.~Raitses, and N.~J. Fisch.
\newblock {Electron cross-field transport in a low power cylindrical Hall
  thruster}.
\newblock {\em Phys. Plasmas}, 11(4922), 2004.

\bibitem{Syd-instability}
D.~Sydorenko, I.~D. Kaganovich, Y.~Raitses, and A.~I. Smolyakov.
\newblock {Breakdown of a Space Charge Limited Regime of a Sheath in a Weakly
  Collisional Plasma Bounded by Walls with Secondary Electron Emission}.
\newblock {\em Phys. Rev. Let.}, 103(145004), 2009.

\bibitem{Pustovitov2000}
V.~D. Pustovitov.
\newblock {\em Theoretical Principles of the Plasma-Equilibrium Control in
  Stellarators}, pages 1--201.
\newblock Springer US, Boston, MA, 2000.

\end{thebibliography}

\end{document}